\documentclass[final,5p,times,twocolumn]{elsarticle}
\usepackage{graphics}
\usepackage{graphicx}
\usepackage{txfonts}
\usepackage{epsfig}
\usepackage{bm}   
\usepackage{dcolumn}
\usepackage{rotate}
\usepackage{amsfonts}
\usepackage{longtable}
\usepackage{latexsym,amssymb,rotating,wrapfig}
\usepackage{footmisc,booktabs}
\usepackage{xcolor,colortbl}
\usepackage{multirow} 
\definecolor{Gray}{gray}{0.85}
\definecolor{LightCyan}{rgb}{0.88,1,1}
\usepackage{amssymb}

\journal{Physics Letters B}

\begin{document}

\begin{frontmatter}

\title{Nuclear polarization effects in Coulomb excitation studies}

\author{J.N. Orce}

\address{Department of Physics, University of the Western Cape, P/B X17, Bellville, ZA-7535 South Africa}
\ead{jnorce@uwc.ac.za}
\ead[url]{http://www.pa.uky.edu/$\sim$jnorce}


\begin{abstract}

New polarization potentials have been determined based on: 1) the latest photo-neutron cross section 
evaluation and a missing factor of two in previous work, and 2) the mass dependency of the symmetry energy, $a_{sym}(A)$. 
The magnitude of the first one is 54\% stronger than the currently accepted polarization potential. 
The second one opens up the possibility for a parameter-free polarization potential. 
Both polarization potentials are essentially the same for heavy nuclei.  
The polarization effect on quadrupole collectivity is more substantial than previously assumed for light nuclei. 
Particular  cases are discussed where long-standing discrepancies between high-precision Coulomb-excitation and lifetime measurements 
still remain.  A solution to the long-standing discrepancy between 
$B(E2; 0^+_1\rightarrow 2^+_1)$ values determined  in $^{18}$O by several 
Coulomb-excitation studies and a high-precision lifetime measurement is provided in favor of the latter.
Polarization effects in light nuclei also influence the determination of spectroscopic quadrupole moments in Coulomb-excitation measurements. 
The hindrance of polarizability  observed in the photo-neutron cross section for single-closed shell nuclei 
is calculated to have a negligible effect on quadrupole collectivity, within the existing experimental uncertainties. 
\end{abstract}

\begin{keyword}
Photo-absorption cross section \sep $E1$ polarizability \sep reduced transition probability \sep  spectroscopic quadupole moment

\end{keyword}

\end{frontmatter}

%
%
%

Virtual excitations are responsible for the polarization of atoms and molecules 
and give rise to the well-known van der Waals forces between two neutral atoms or molecules, 
which are far enough apart for the ovelap between the wave functions to be neglected~\cite{mott}. 
In nuclei, electric-dipole virtual excitations via high-lying states in the 
giant dipole resonance, {\small GDR}~\cite{GDRreview}, can also polarize 
the ground and excited states of nuclei~\cite{eichler,alder,levinger2}. 
This polarization phenomenon is the so-called \emph{$E1$ polarizability} and 
is directly related to the static nuclear polarizability, $\alpha$.\\

The ability for a nucleus to be polarized is driven by the dynamics of the {\small GDR}, i.e., 
the inter-penetrating motion of proton and neutron fluids out of phase~\cite{migdal2}. 
This motion results in the nuclear symmetry energy, 
$a_{sym}(A)(\rho_n-\rho_p)^2/\rho$, acting as a restoring force~\cite{migdal2,migdal}. 
The nuclear symmetry energy parameter, $a_{sym}(A)$, 
is key to undertanding the elusive equation of state of neutron-rich matter, which impacts 
three-nucleon forces~\cite{hebeler}, neutron skins~\cite{nskin,nskin2}, neutron stars 
and supernova cores~\cite{neutronstars,latimer,latimer2,pearson}. 
The hydrodynamic model connects  $\alpha$ and  $a_{sym}(A)$ by~\cite{migdal,levinger,orce1}, 
\begin{equation}
\alpha=\frac{e^2R^2A}{40~a_{sym}(A)} \mbox{fm}^3. 
\label{eq:sigma-2}
\end{equation}

In nuclear reactions, the induction of an electric dipole moment {\bf p} in the nucleus can be generated by 
the time-dependent electric field {\bf E} of the partner. 
The nuclear polarizability $\alpha=\frac{\mathbf{p}}{\mathbf{E}}$ can also be determined using second-order perturbation theory, 
\begin{eqnarray}
\alpha&=&2e^2\sum_n \frac{\langle i\parallel\hat{E1}\parallel n\rangle \langle n\parallel\hat{E1}\parallel 
i\rangle}{E_{_{\gamma}}}=\frac{\hbar c}{2\pi^2}\sigma_{_{-2}},  
\label{eq:polar} 
\end{eqnarray}
where  $\sigma_{_{-2}}$ is the $(-2)$ moment of the total electric-dipole 
photo-absorption cross section~\cite{levinger2,migdal3}.
This sum rule indicates that large $E1$ matrix elements via virtual excitations of the 
\emph{\small GDR}~\cite{GDRenergy} 
may polarize the shape of the  ground  state $|i\rangle$. 

In Coulomb-excitation studies, two-step processes of the type $|i\rangle \rightarrow |n\rangle \rightarrow |f\rangle$ 
(e.g., $0^+_1 \rightarrow 1^-_{_{GDR}} \rightarrow 2^+_1$) can  affect 
the extracted reduced transition probability, i.e., the $B(E2)$ values, and  
the sign and magnitude of the spectroscopic quadrupole moment, {\small $Q_{_S}$}, of the final excited state $|f\rangle$~\cite{hausser2}.  
A final state  with $J^{\pi}=2^+$ is assumed hereafter for excited states.  
Both the \emph{$E1$ polarizability} and the \emph{reorientation effect}, {\small $RE$,} are second-order effects 
in Coulomb-excitation theory~\cite{eichler,deBoerEichler,hausser0,alder}. 
The {\small RE} generates a time-dependent hyperfine splitting of nuclear levels~\cite{hausser0},  
which can be used to determine the nuclear charge distribution in the laboratory frame~\cite{deBoerEichler,hausser0}, 
i.e.,  {\small $Q_{_S}$}, for states with angular momentum $J\neq 0,\frac{1}{2}$.
The angular distribution of the de-excited $\gamma$-rays as a function of scattering angle 
may be enhanced ({\small $Q_{_S}(2_1^+)>0$}) or inhibited ({\small $Q_{_S}(2_1^+)<0$}), 
hence providing a spectroscopic probe for a measurement of {\small $Q_{_S}$}.\\

\begin{figure}[!h]
\begin{center}
\includegraphics[width=8.2cm,height=7.5cm,angle=-0]{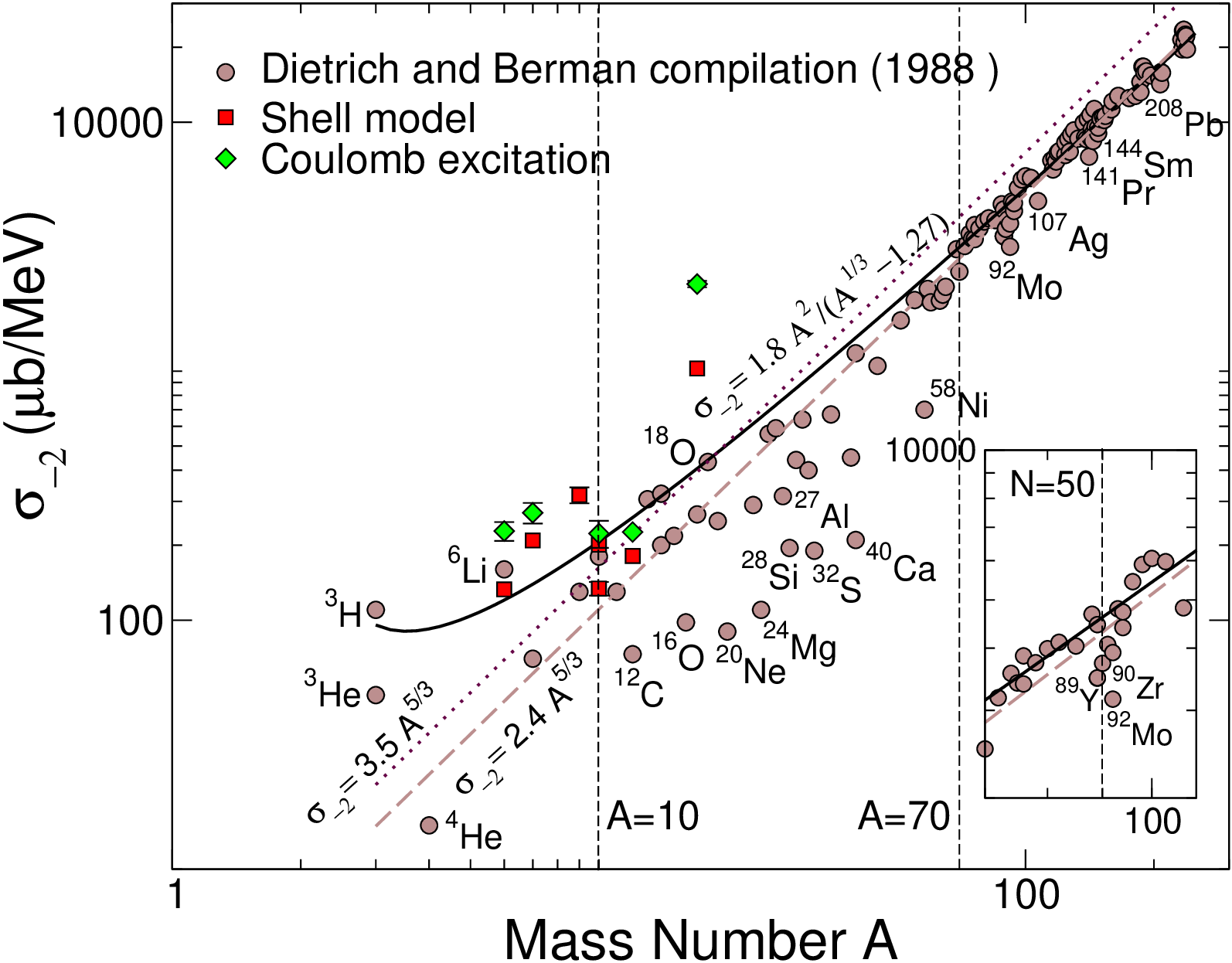} 
\caption{(Color online) The $\sigma_{_{-2}}$ data $vs$ A  on a log-log scale from the 1988 photo-neutron 
cross-section evaluation (circles)~\cite{atlas}, shell model (squares)~\cite{barker,barker2,orce} and Coulomb-excitation (diamonds) 
\cite{hausser2,hausser3,6li,10B_vermeer,bamberger} studies.
For comparison, Eqs.~\ref{eq:3p5} (dotted line), \ref{eq:mine} (dashed line) and \ref{eq:mine2} (solid line) are also plotted. 
Shell effects are noticeable for the $N=50$ (inset), $N=82$ and $N=126$ isotones.}
\label{fig:sigmasym}
\end{center}
\end{figure}

The polarization potential $V_{pol}$ generated by the $E1$ polarizability is proportional to $\alpha$, 
and reduces the  effective quadrupole interaction $V_{_{eff}}(t)$ in the following manner~\cite{hausser2},
\begin{eqnarray}
 V_{_{eff}}(t) &=&  ~V_{_0}(t)~ \left(~1-V_{pol}(t)\right) \label{eq:pol} \\ 
 &=&  ~V_{_0}(t) \left(1-z\frac{a}{r(t)}\right)\nonumber.
\end{eqnarray}
For the case of projectile 
excitation,\footnote{Similarly, for the case of target excitation, $z \approx 0.005 \kappa \frac{E_p A_t}{Z_t^2(1+A_p/A_t)}$.}
$z$ is  given by Alder and Winther (appendix J)~\cite{alder},  
\begin{equation}
z= \frac{10Z_t\alpha}{3Z_pR^2a} \approx 0.005 \kappa ~\frac{E_p A_p}{Z_p^2(1+A_p/A_t)}, 
\label{eq:0005}
\end{equation}
with $E_p$ the kinetic energy (in MeV) in the laboratory frame, 
$a$ the half-distance of closest approach in a head-on collision, $r(t)$ the magnitude of the projectile-target position vector, 
and $\alpha$ is given by combining\footnote{Equation~\ref{eq:polar} is in better agreement with recent measurements of $\alpha(^{208}\mbox{Pb})=18.9(13)$ fm$^3$ 
using polarized protons~\cite{tamii}. From the empirical value of $\sigma_{_{-2}}=1.59$ fm$^2$/MeV~\cite{atlas} 
and Eq.~\ref{eq:polar}, $\alpha(^{208}\mbox{Pb})=15.9$ fm$^3$.} $\alpha=\frac{\hbar c}{4\pi^2}\sigma_{_{-2}}$~\cite{alder} and Levinger's 
empirical formula~\cite{levinger},
\begin{equation}
\sigma_{_{-2}}=3.5\kappa\times 10^{-4} A^{5/3} ~\mbox{fm$^2$/MeV},  
\label{eq:3p5}
\end{equation}
where  a constant value of $a_{sym}=23$ MeV is assumed, and 
the polarizability parameter $\kappa$ is the ratio of the observed {\small GDR} effect for ground states 
(circles in Fig.~\ref{fig:sigmasym}) 
to that predicted by the hydrodynamic model~\cite{migdal,levinger}. A value of $\kappa=1$  is broadly accepted since 1957 for 
$A\geq 20$ nuclei~\cite{levinger}. 
The general assumption in Coulomb-excitation codes~\cite{gosia,tomdrake} is that the $E1$ polarizability effect 
for excited states remains the same as the one determined by Levinger for ground states~\cite{levinger}. 
Equation~\ref{eq:3p5} (for $\kappa=1$) is plotted (dotted line) in Fig.~\ref{fig:sigmasym}. 

The factor of 0.005 in front of Eq.~\ref{eq:0005}, the so-called $E1$ polarization parameter,
is the default value (for $\kappa=1$)  used in modern Coulomb-excitation codes such as 
{\small GOSIA}~\cite{gosia} and  Winther$-$de Boer~\cite{tomdrake}.  
Curiously, H\"ausser~\cite{hausser0} provides an equality to Eq.~\ref{eq:0005} with the actual 
factor of 0.0056 given by Nakai and Winther's private communication~\cite{hausser3,winther}, but invoking 
the right-hand side of Eq.~\ref{eq:polar}. 
Instead, the factor of 0.005 in Eq.~\ref{eq:0005} was deduced by Alder and Winther using 
$\alpha=\frac{\hbar c}{4\pi^2}\sigma_{_{-2}}$ from Ref.~\cite{levinger2}. 
Nevertheless, Eqs. (1.6), (1.7), (1.8), (1.9) and (1.24) in Levinger's monograph~\cite{levinger2} yield 
Eq.~\ref{eq:polar}, in agreement with Migdal~\cite{migdal} and Levinger himself~\cite{levinger}. 
With the correct use of Eq.~\ref{eq:polar}~\cite{mott,merzbacher,levinger,levinger2,orce1,hausser0}, 
the resulting polarization potential is two times larger,  
\begin{equation}
z=  0.0112~\kappa ~\frac{E_p A_p}{Z_p^2(1+A_p/A_t)}. 
\label{eq:0011}
\end{equation}
This enhancement results in a substantial reduction of the effective quadrupole interaction. 
With the  polarization potential in Eq.~\ref{eq:0011}, 
calculations with the semi-classical coupled-channel Coulomb excitation code {\small GOSIA}~\cite{gosia} 
generally yield (for $\kappa=1$) an increase of approximately 6\% with respect to previously extracted $B(E2)$ values.

This long-standing misunderstanding is fortunately mitigated by a new empirical formula for $\sigma_{_{-2}}$~\cite{orce1} 
determined from the 1988 photo-neutron cross-section evaluation using monoenergetic photons~\cite{atlas},
\begin{equation}
\sigma_{-2}=2.4k A^{5/3} \mu\mbox{b/MeV}. 
\label{eq:mine}
\end{equation}
This equation is plotted (dashed line) in Fig.~\ref{fig:sigmasym}. The overall modification of the polarization potential is, therefore, 
\begin{equation}
z = 0.0077~\kappa~\frac{E_p A_p}{Z_p^2(1+A_p/A_t)},
\label{eq:0038}
\end{equation}
which gives an overall increase of 54\%   with respect to the polarization potential in Eq.~\ref{eq:0005}, 
and  an increase of approximately 3\% (for $\kappa=1$) in previously extracted $B(E2)$ values. \\

\begin{table*}[!ht]
\centering
\begin{tabular}{|c|c|c|c|c|c|c|c|c|} 
\cline{4-9}
\multicolumn{3}{c|}{} & \multicolumn{4}{c|}{$\sigma_{_{-2}}=3.5\kappa A^{5/3}$ $\mu$b/MeV} & \multicolumn{2}{c|}{$\sigma_{_{-2}}=2.4\kappa A^{5/3}$ $\mu$b/MeV} \\
\hline \hline
Nucleus     &  E$_x$(MeV) &    $J^{\pi}$   & $\kappa_{_{SM}}$ &   Ref.         & $\kappa_{_{Coulex}}$ &  Ref. & $\kappa_{_{SM}}$  & $\kappa_{_{Coulex}}$  \\ \hline \hline
$^6$Li      &   2.186     &    3$^+_1$     &  1.9           & \cite{barker}  &  3.3(7)              & \cite{6li}                                                    & 2.8               &  4.8(10)    \\ \hline
$^7$Li      &   0.478     &    1/2$^-_1$   &  2.3           & \cite{barker}  &  3.0(7)              & \cite{hausser2,bamberger,hausser3,smilansky,7Li_Vermeer}      & 3.4               &  4.4(10)    \\ \hline
$^9$Be      &   0         &    3/2$^-_1$   &  2.3(4)        & \cite{orce}    &                      &                                                               & 3.4(6)            &             \\ \hline
$^{10}$B    &   0.718     &    1$^+_1$     &  1.2           & \cite{barker}  &  1.4(4)*             & \cite{10B_vermeer}                                            & 1.8               &  2.0(6)*    \\ \hline
$^{10}$Be   &   3.368     &    2$^+_1$     &  0.8(2)        & \cite{orce}    &                      &                                                               & 1.2(3)            &             \\ \hline
$^{12}$C    &   4.439     &    2$^+_1$     &  0.8           & \cite{barker}  &  1.0*                & \cite{12C_vermeer}                                            & 1.2               &  1.5*       \\ \hline
$^{17}$O    &   0.871     &    1/2$^+_1$   &  2.6           & \cite{barker2} &  5.7(4)              & \cite{17O}                                                    & 3.8               &  8.3(6)     \\ \hline
\hline
\end{tabular}
\caption{Polarizability parameters $\kappa$ determined from Coulomb-excitation measurements 
(Coulex)~\cite{6li,hausser2,10B_vermeer,bamberger,hausser3} and shell-model calculations~\cite{barker,barker2,orce} using the empirical formulas of Levinger [Eq.~\ref{eq:3p5}] and Orce [Eq.~\ref{eq:mine}]. 
An asterisk indicates that crude estimations were  made to determine $\kappa$~\cite{10B_vermeer,12C_vermeer}. 
\label{tab:k}}
\end{table*}

A broader description of the polarization potential naturally arises from the inclusion of $a_{sym}(A)$ in Eq.~\ref{eq:sigma-2}. 
The mass dependency of $a_{sym}(A)$ has recently been determined by Tian and collaborators~\cite{tian} from 
a global fit to the binding energies of 
isobaric nuclei with $A\geq10$, extracted from the 2012 atomic mass evaluation~\cite{audi}, 
\begin{equation}
 a_{sym}(A)=28.32\left(1-1.27A^{-1/3}\right),
 \label{eq:asym}
\end{equation}
where $S_v\approx 28.32$ MeV is the bulk symmetry energy coefficient 
and $\frac{S_s}{S_v}\approx 1.27$ the surface-to-volume ratio~\cite{tian}. 
Equation~\ref{eq:asym}  includes  Coulomb energy and shell corrections and, more importantly, 
opens up the exciting prospect of generic equations for $\alpha$ and $\sigma_{_{-2}}$, 
\begin{equation}
 \alpha(A)=\frac{1.83\times10^{-3} ~A^2}{A^{1/3}-1.27}~\mbox{fm}^3,  
 \label{eq:mine3}
\end{equation}
\begin{equation}
\sigma_{_{-2}}(A)=\frac{1.83 ~A^{2}}{A^{1/3}-1.27} ~\mu\mbox{b/MeV}, 
\label{eq:mine2}
\end{equation}
which may not require introducing the empirical parameter $\kappa$~\cite{orce1}. 
Equation~\ref{eq:mine2} is plotted  in Fig.~\ref{fig:sigmasym} (solid line) for $A\geq3$ nuclides, 
and provides an explanation for the enhancement of $\sigma_{_{-2}}$ observed in light nuclei. 
However, Eq.~\ref{eq:mine2} becomes negative at $A=2$. 
The fact that the {\small GDR} has been observed 
for all nuclei except for the deuteron may support Eq.~\ref{eq:mine2} for $A\geq3$. 
The predicted curve nicely merges with the 
data for $A\gtrsim70$, in agreement with the dominant photo-neutron cross sections in heavy nuclei~\cite{orce1}. 

\begin{figure}[!h]
\begin{center}
\includegraphics[width=6.5cm,height=6.cm,angle=-0]{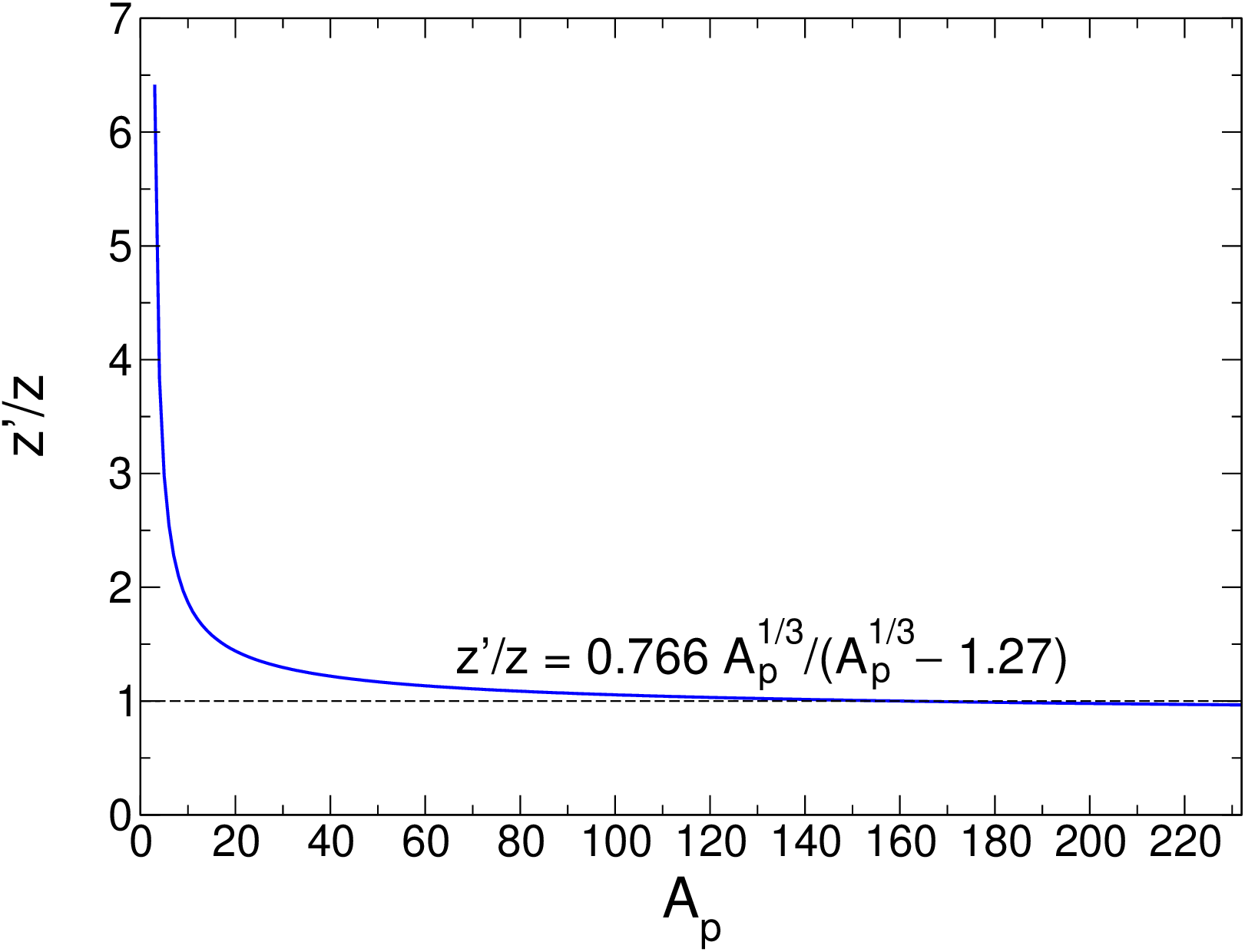} 
\caption{(Color online) Plot of $z^{\prime}/z$ (as given in Eqs.~\ref{eq:0038} and ~\ref{eq:0059})  $vs$ A for $\kappa=1$.}
\label{fig:zprimeoverz}
\end{center}
\end{figure}

Finally, a $\kappa$-free polarization potential $z^{\prime}$ can be obtained inserting Eq.~\ref{eq:mine3} 
on the left-hand side of Eq.~\ref{eq:0005},
\begin{equation}
z^{\prime} = 0.0059 ~\frac{E_p}{Z_p^2(1+A_p/A_t)}\frac{A_p^{4/3}}{A_p^{1/3}-1.27}.
\label{eq:0059}
\end{equation}
A plot of $z^{\prime}/z$ for $\kappa=1$ as a function of mass number $A$  is shown in Fig.~\ref{fig:zprimeoverz} for 
$A\geq3$ nuclides. 
It is evident that the mass dependence of $a_{sym}(A)$ gives rise to a larger polarization potential $z^{\prime}$ 
and an additional enhancement of quadrupole collectivity for light nuclei. It is reassuring to find out that both polarization 
potentials converge as $A$ increases. 
The hydrodynamic model still remains to be  tested for light nuclei. 



The polarizability effect of Eq.~\ref{eq:0038} on matrix elements extracted from Coulomb-excitation measurements 
may additionally  be enhanced or inhibited depending on the $\kappa$ value. 
As shown in Fig.~\ref{fig:sigmasym}, 
deviations from the smooth trend in Eq.~\ref{eq:0038} are observed for 
$A=4n$, $T_{_Z}=0$ nuclei ($\kappa<1$), loosely-bound light nuclei with $A<20$ ($\kappa>1$), 
and single-closed shell nuclei with $N=50$, $N=82$ and $N=126$ ($\kappa<1$). 
The drop of $\sigma_{_{-2}}$ below $A\lesssim50$, including $^{58}$Ni, is due to the 
missing $(\gamma,p)$ contributions in the Dietrich and Berman evaluation~\cite{atlas}. The enhancement 
of $\sigma_{_{-2}}$ for loosely-bound light nuclei is due to the smaller $a_{sym}(A)$ values~\cite{orce1}. 
The hindrance of polarizability for single-closed shell nuclei (e.g., as shown in the inset of Fig.~\ref{fig:sigmasym} 
for the $N=50$ isotones) 
is due to the resistance of more tightly-bound spherical nuclei 
to be polarized. 

Polarizability parameters for excited states in light nuclei have been 
determined in favorable Coulomb-excitation measurements~\cite{6li,hausser2,10B_vermeer,bamberger,hausser3}, 
where the validity of the hydrodynamic model [Eq.~\ref{eq:3p5}] was assumed~\cite{6li,bazhanov,hausser2,7Li_Vermeer,smilansky,10B_vermeer,12C_vermeer,bamberger,17O}. 
The reason for the scarcity of experimental work 
is that there are a few good candidates where 
$\kappa$ can be extracted independently of {\small $Q_{_S}$}. 
The Coulomb-excitation data do not follow the smooth trend predicted by Eq.~\ref{eq:mine2}, particularly for $^{17}$O. 
However, these results were generally extracted from particle spectra and modern particle-$\gamma$ coincidence measurements 
will be very useful. 
In addition, shell model calculations~\cite{barker,barker2,orce} have been performed to compute $\kappa$,  
assuming that all the $E1$ strength from the ground state is concentrated at $E_{_{GDR}}$~\cite{barker,barker2}, 
and with the no-core shell model~\cite{orce}. 
The  results are shown in Fig.~\ref{fig:sigmasym} and listed in Table~\ref{tab:k}, and 
have been renormalised according to Eq.~\ref{eq:mine} (last two columns in Table~\ref{tab:k}).  
Large $\kappa$ parameters are generally determined for loosely-bound light 
nuclei~\cite{hausser2,10B_vermeer,12C_vermeer,6li,bazhanov}. 
It is relevant to investigate deviations from $\kappa\neq1$ using Eqs.~\ref{eq:0038} and \ref{eq:0059} together 
with the available information on $\kappa$ obtained from the photo-absortion cross sections, Coulomb-excitation 
measurements and shell-model calculations.

Light nuclei present large polarizability parameters of $\kappa>1$, which can enhance quadrupole collectivity. 
A curious case is the large $\kappa=8.3(6)$ determined in $^{17}$O~\cite{17O}. 
As suggested by  Kuehner and co-workers~\cite{17O}, a slightly larger than $\kappa>1$ value  in $^{18}$O 
would explain the long-standing $\approx$12\% discrepancy between the smaller $B(E2; 0^+_1\rightarrow 2^+_1)$ 
determined from seven Coulomb excitation measurements, 0.00421(9) e$^2$b$^2$~\cite{raman},  
and the larger one extracted from a high-precision lifetime measurement, 0.00476(11) e$^2$b$^2$, 
determined by Ball and co-workers in 1982 by fitting the Doppler-broadened $\gamma$-ray lineshapes~\cite{ball}. 
The latter value nicely agrees with a similar lifetime measurement by Hermans and co-workers~\cite{hermans}. 
A half-way $B(E2; 0^+_1\rightarrow 2^+_1)=0.00448(13)$ e$^2$b$^2$ is determined from an 
inelastic electron scattering measurement by Norum and collaborators in 1982~\cite{norum_eep}.

Using Eq.~\ref{eq:mine}, a value of $\kappa\approx1.8$ for the ground state of $^{18}$O is determined 
from $\sigma_{_{-2}}=547~\mu$b/MeV~\cite{woodworth_totalphoto}. This $\sigma_{_{-2}}$ value was determined from 
total photonuclear cross sections, which included $\sigma(\gamma,p)$, $\sigma(\gamma,n)$ + $\sigma(\gamma,np)$, and $\sigma(\gamma, 2n)$, 
and was integrated from threshold to 42 MeV. 
As shown in Fig.~\ref{fig:zprimeoverz}, 
this relatively large $\kappa$ value is in agreement with the larger $z^{\prime}/z\approx1.5$ observed for $A_p=18$.  
Using $\kappa=1.8$ for the 2$^+_1$ state in $^{18}$O in  Eq.~\ref{eq:0038}, a {\small GOSIA} calculation 
of  $^{18}$O beams at a safe energy of 90 MeV scattered off a $^{208}$Pb target with a [30$^{\circ}$,60$^{\circ}$] angular coverage
yields an increase of $\approx$10\% in the $B(E2; 0^+_1\rightarrow 2^+_1)$ value relative 
to the one given by  Eq.~\ref{eq:0005}. 
This relative increase is independent of the $\langle 2^+_1\mid\mid \hat{E2} \mid\mid 0^+_1\rangle$ and 
$\langle 2^+_1\mid\mid \hat{E2} \mid\mid 2^+_1\rangle$ matrix elements. 
As shown in Fig.~\ref{fig:sigmasym}  and Table~\ref{tab:k}, similar or larger enhancements of  
$B(E2)$ values can be expected for other light nuclei. 

%

For light nuclei, the enhancement of polarizability will also shift the sign and magnitude of {\small $Q_{_S}$}~\cite{hausser2,orce1} towards the oblate side. For well-deformed nuclei, the quadrupole interaction dominates and low-lying  states are 
more relevant to the 2$^+_1$ excitation cross section than the $E1$ polarizability~\cite{hausser0,deBoerEichler}.
Because of the $A_p/Z_p^2$ dependence,  
a large reduction of the excitation cross section occurs for light nuclei, where the effect of the 
$E1$ polarizability may exceed the reorientation effect~\cite{6li,bamberger,hausser2,hausser3}. 
No photo-absorption cross section information is available for radioactive nuclei. 
However, the renormalised value of $\kappa=1.2$  (see Table \ref{tab:k}) 
calculated with the 
no-core shell model for the 2$^+_1$ state in $^{10}$Be~\cite{orce}
yields a shift of approximately  $+0.05$ eb in the $Q_{_S}(2^+_1)$ value~\cite{orce3}. This shift aligns better with 
the no-core shell model calculations of $Q_{_S}(2^+_1)$ in $^{10}$Be~\cite{orce}. 

The hindrance of polarizability observed in Fig.~\ref{fig:sigmasym} for single-closed shell nuclei 
has not  previously been addressed and deserves investigation for prominent cases. 
The inset of Fig.~\ref{fig:sigmasym} shows a sharp drop of  $\sigma_{_{-2}}$ values for the 
$N=50$ isotones. These shell effects arise from the resistance 
of single-closed shell nuclei to be polarized and are the result of the more bound spherical density 
distribution caused by the short-range pairing interaction~\cite{rowewood}.
Considering $\kappa=0.7$ for the 2$^+_1$ state in $^{92}$Mo, as given by $\sigma_{_{-2}}=3160$ $\mu$b/MeV~\cite{atlas} and Eq.~\ref{eq:mine}, 
a reduction of 3\% for the  $B(E2; 2^+_1 \rightarrow 0^+_1)$ value is calculated  using the polarization potential in 
Eq.~\ref{eq:0038}. This reduction  is not significant within the existing 6\% uncertainty~\cite{nndc}. 
Because of the 30\% reduction in  Eq.~\ref{eq:0038}, this result  agrees with 
the $B(E2)$ value extracted using the polarization potential in Eq.~\ref{eq:0005}.
These {\small GOSIA} calculations assume a $^{92}$Mo beam at a safe energy 
of 480 MeV scattering off a $^{208}$Pb target with a [30$^{\circ}$,60$^{\circ}$] angular coverage.

\begin{figure}[!h]
\begin{center}
\includegraphics[width=4.3cm,height=5.cm,angle=-0]{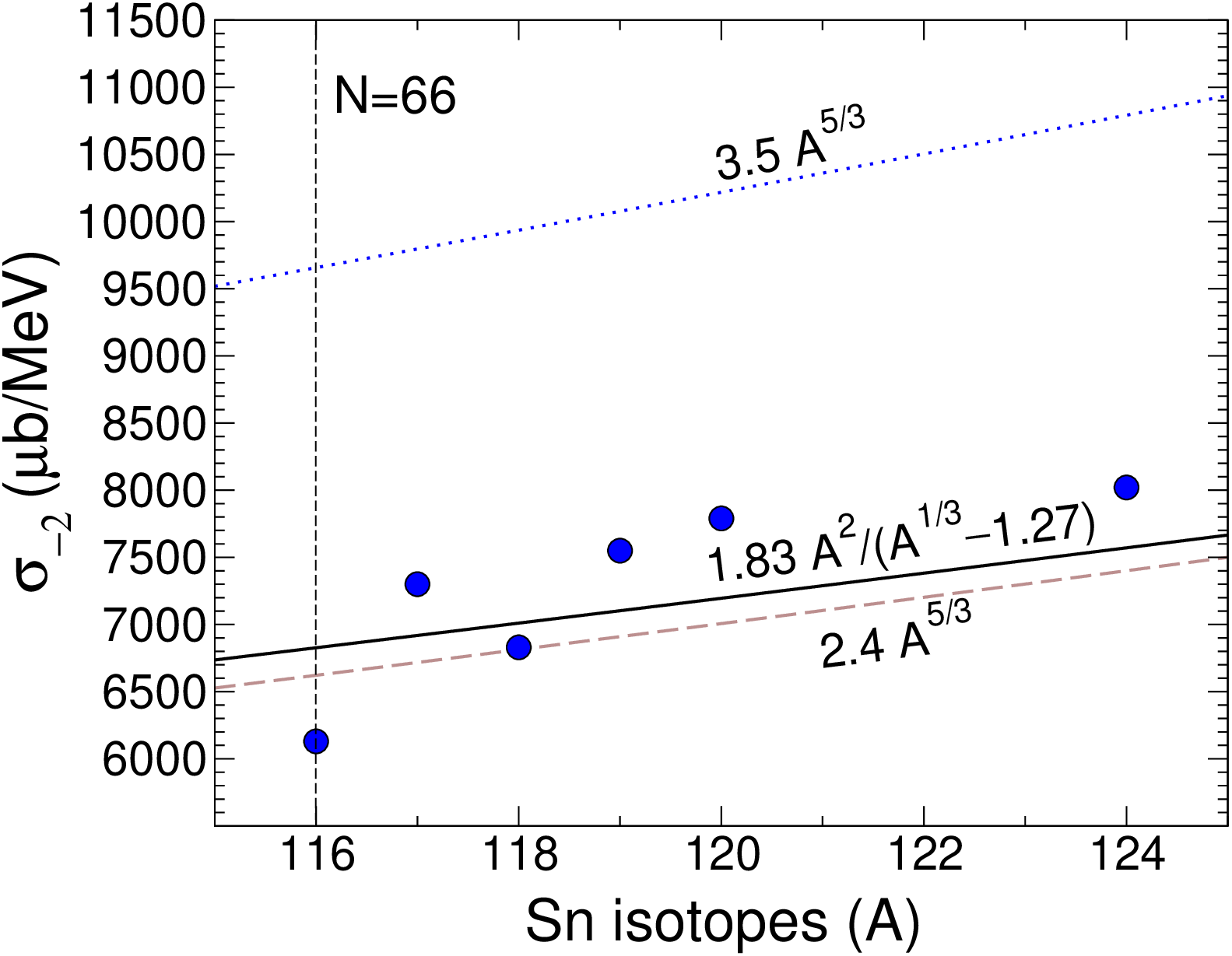}
\hspace{1.3mm}
\includegraphics[width=3.8cm,height=5.cm,angle=-0]{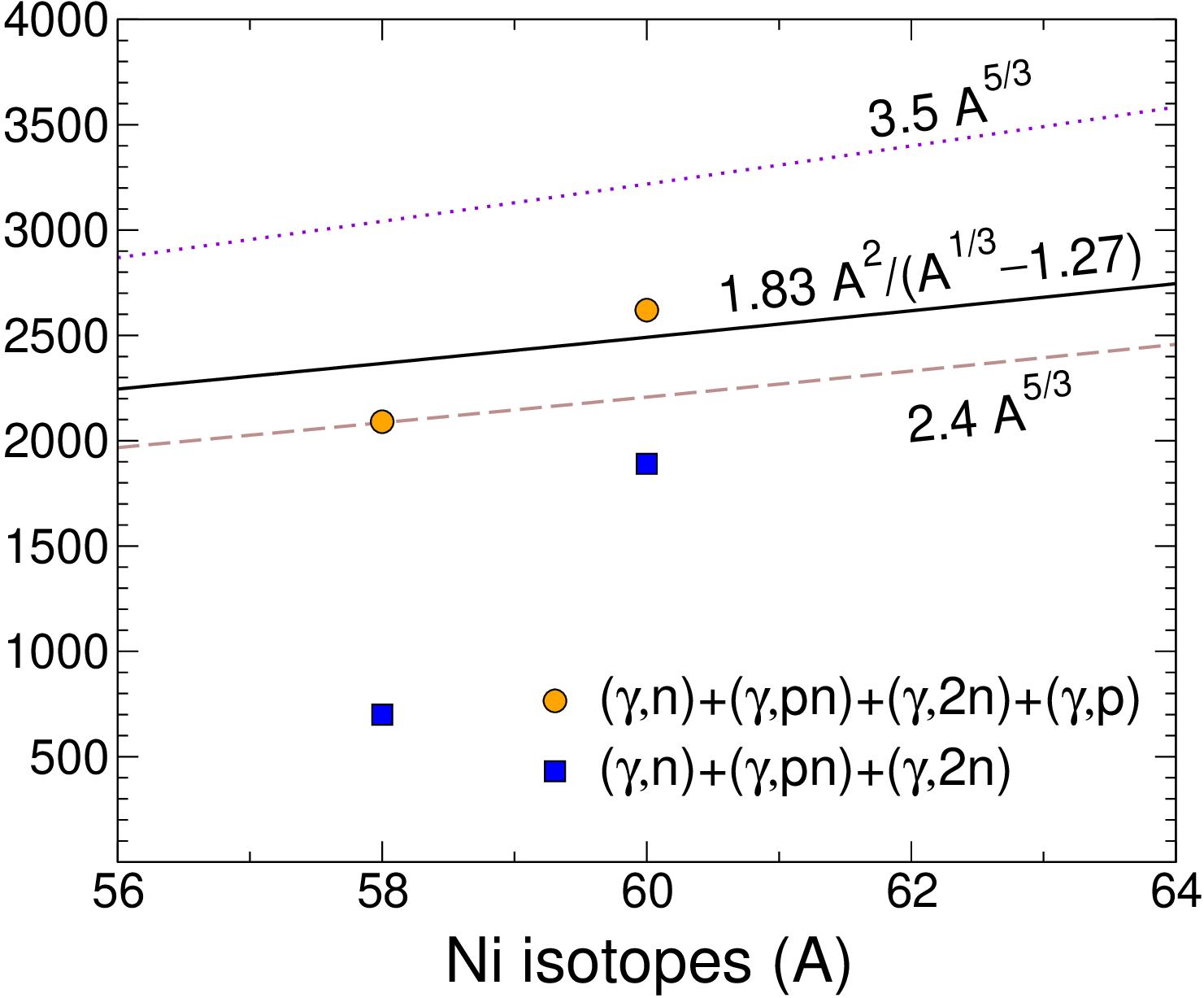} 
\caption{(Color online) Empirical  $\sigma_{_{-2}}$ values available for the Sn (left) and Ni (right) isotopes~\cite{116sn}. 
For comparison, Eqs.~\ref{eq:3p5}, (dotted line),~\ref{eq:mine} (dashed line) and \ref{eq:mine2} (solid line) 
are plotted.}
\label{fig:Sn_sigma}
\end{center}
\end{figure}

Additional shell effects are shown in Fig.~\ref{fig:Sn_sigma} (left panel) for the Sn isotopes. 
A  sudden drop of $\sigma_{_{-2}}=6130$ $\mu$b/MeV for $^{116}$Sn~\cite{116sn} is observed relative to the heavier tin isotopes. 
Although this small drop ($\kappa=0.93$ using Eq.~\ref{eq:mine}) may hint a weak subshell gap at $N=66$, 
it actually increases by 
2\% the $B(E2; 2^+_1\rightarrow 0^+_1)$ value determined using Levinger's formula. 
{\small GOSIA} calculations 
of $^{116}$Sn beams at 580 MeV scattered off a $^{208}$Pb target with a [30$^{\circ}$,60$^{\circ}$]  angular coverage 
cannot accommodate, even using a much smaller $\kappa$ value, the 20\% discrepancy 
between Coulomb-excitation measurements~\cite{raman,allmond2} and the high-precision lifetime measurement by Jungclaus and 
co-workers~\cite{andrea}. 

Another peculiar case concerns the smaller $B(E2; 2^+_1\rightarrow 0_1^+)$  value extracted from a 
high-precision lifetime measurement by Kenn {\it et al.}~\cite{kenn}, which is several standard deviations from the 
2001 evaluation of Raman {\it et al.}~\cite{raman}. 
The right panel of Fig.~\ref{fig:sigmasym} shows a  drop of $\sigma_{_{-2}}$ in $^{58}$Ni. 
However, once the $(\gamma,p)$ contribution is added, $\sigma(\gamma,p)\approx3\sigma(\gamma,n)$~\cite{58ni}, $\sigma_{_{-2}}$ nicely aligns 
with Eq.~\ref{eq:mine} (for $\kappa=1$). The $B(E2; 2^+_1\rightarrow 0_1^+)$ value extracted from Coulomb-excitation 
measurements is only enhanced by approximately 3\%, according to Eq.~\ref{eq:0038}, and cannot accomodate the 
discrepancy between Coulomb-excitation and lifetime measurements. In fact, 
the extracted $E2$ strengths from lifetime measurements using the inelastic neutron scattering reaction~\cite{orce2}
and recent high-precision Coulomb excitation studies by Allmond and collaborators~\cite{mitch} are both in 
agreement with the 2001 evaluation of Raman {\it et al.}~\cite{raman}.

In conclusion, new polarization potentials have been determined in this work [Eqs.~\ref{eq:0038} and \ref{eq:0059}]. 
The polarization potential $z$ in Eq.~\ref{eq:0038} arises from the latest photo-neutron cross section evaluation and 
a missing factor of two in previous work, and yields an overall increase of 54\% with respect to the 
broadly accepted polarization potential [Eq.~\ref{eq:0005}]. 
The second one, $z^{\prime}$ in Eq.~\ref{eq:0059}, opens up the possibility for 
a $\kappa$-free polarization potential based on the mass dependence of  $a_{sym}(A)$. 
A plot of $z^{\prime}/z$ vs. $A$ shows that the two potentials are essentially the same for heavy nuclei. 
Moreover, $z^{\prime}$  explains why light nuclei present an enhanced quadrupole collectivity. 
Polarization effects in light nuclei affect the determination of $B(E2)$ and $Q_{_S}$ values in 
Coulomb-excitation studies more than previously assumed. 
In particular, a value of $\kappa=1.5$ in $^{18}$O deduced from the latest photo-neutron cross section evaluation, 
and independently by $z^{\prime}/z$ at $A=18$, remarkably explains the 
long-standing discrepancy between the $B(E2; 0_1^+\rightarrow 2_1^+)$ values determined from seven 
Coulomb-excition studies and one high-precision lifetime measurement. 
A shift of approximately $+0.05$ eb is also 
determined for $Q_{_S}(2^+_1)$ in $^{10}$Be, and results in a better agreement with no-core shell model 
calculations. 
In general, the hindrance of polarization observed in the photo-neutron cross section data 
has a negligible effect in quadrupole collectivity, within the existing uncertainties.

Finally, it is important to remark the scarce information available concerning the \emph{E1 polarizability}. 
To date, our knowledge on how nuclei polarize mainly arises from the photo-absorption cross-section 
data of ground states in stable nuclei. These polarization effects are assumed to behave similarly for excited 
states populated in the Coulomb excitation of stable and radioactive nuclei. 
As suggested by Eichler~\cite{eichler} and H\"ausser~\cite{hausser0}, modern Coulomb-excitation 
particle-$\gamma$ measurements at different bombarding energies and covering large scattering 
angular ranges present a potential spectroscopic probe to disentangle 
the \emph{E1 polarizability} and $Q_{_S}$. These measurements 
may provide valuable information on deviations from the hydrodynamic model and an alternative probe  
to constrain $a_{sym}(A)$. Additional experimental and theoretical work are clearly demanded. 


The author acknowledges fruitful physics discussions with G. C. Ball, B. A. Brown,  S. Triambak, D. H. Wilkinson, J. L. Wood and S. W. Yates. 
This work was supported by the South African National Research Foundation (NRF) under Grant 93500.

\end{document}